\documentstyle[12pt,psfig]{article} 
\makeatletter \def\@cite#1#2{{[{#1}]\if@tempswa\typeout {IJCGA
warning: optional citation argument ignored: `#2'} \fi}}

\newcount\@tempcntc
\def\@citex[#1]#2{\if@filesw\immediate\write\@auxout{\string\citation{#2}}\fi
  \@tempcnta\z@\@tempcntb\m@ne\def\@citea{}\@cite{\@for\@citeb:=#2\do
    {\@ifundefined
       {b@\@citeb}{\@citeo\@tempcntb\m@ne\@citea\def\@citea{,}{\bf ?}\@warning
       {Citation `\@citeb' on page \thepage \space undefined}}%
    {\setbox\z@\hbox{\global\@tempcntc0\csname b@\@citeb\endcsname\relax}%
     \ifnum\@tempcntc=\z@ \@citeo\@tempcntb\m@ne
       \@citea\def\@citea{,}\hbox{\csname b@\@citeb\endcsname}%
     \else
      \advance\@tempcntb\@ne
      \ifnum\@tempcntb=\@tempcntc
      \else\advance\@tempcntb\m@ne\@citeo
      \@tempcnta\@tempcntc\@tempcntb\@tempcntc\fi\fi}}\@citeo}{#1}}
\def\@citeo{\ifnum\@tempcnta>\@tempcntb\else\@citea\def\@citea{,}%
  \ifnum\@tempcnta=\@tempcntb\the\@tempcnta\else
   {\advance\@tempcnta\@ne\ifnum\@tempcnta=\@tempcntb \else 
\def\@citea{--}\fi
    \advance\@tempcnta\m@ne\the\@tempcnta\@citea\the\@tempcntb}\fi\fi}
\makeatother

\def\boxit#1{\leavevmode\thinspace\hbox{\vrule\vtop{\vbox{\hrule%
        \vskip3pt\kern1pt\hbox{\vphantom{\bf/}\thinspace\thinspace%
        {\bf#1}\thinspace\thinspace}}\kern1pt\vskip3pt\hrule}\vrule}%
        \thinspace}
\def\Boxit#1{\noindent\vbox{\hrule\hbox{\vrule\kern3pt\vbox{
\advance\hsize-7pt\vskip-\parskip\kern3pt\bf#1 \hbox{\vrule height0pt
depth\dp\strutbox width0pt} \kern3pt}\kern3pt\vrule}\hrule}}

\newcommand{\cbeta}{\stackrel{\circ}{\beta}}

\newcommand{\Hh}{\lower1.2ex\hbox{$\stackrel{\textstyle
H}{\footnotesize\sim}$}}
\newcommand{\Hho}{\lower1.2ex\hbox{$\stackrel{\textstyle
H_1}{\footnotesize\sim}$}}
\newcommand{\Hhw}{\lower1.2ex\hbox{$\stackrel{\textstyle
H_2}{\footnotesize\sim}$}}
\newcommand{\h}{\lower1.2ex\hbox{$\stackrel{\textstyle
h}{\footnotesize\sim}$}}

\newcommand{\gsim}{\lower.7ex\hbox{$\;\stackrel{\textstyle>}{\sim}\;$}}
\newcommand{\lsim}{\lower.7ex\hbox{$\;\stackrel{\textstyle<}{\sim}\;$}}
\newcommand{\be}{\begin{equation}} \newcommand{\ee}{\end{equation}}
\newcommand{\beq}{\begin{equation}} \newcommand{\eeq}{\end{equation}}
\newcommand{\bea}{\begin{eqnarray}} \newcommand{\eea}{\end{eqnarray}}

\def\simlt{\stackrel{<}{{}_\sim}} \def\simgt{\stackrel{>}{{}_\sim}}

\def\baselinestretch{1}
\begin{document}
\catcode`@=11 \newtoks\@stequation
\def\subequations{\refstepcounter{equation}%
\edef\@savedequation{\the\c@equation}%
\@stequation=\expandafter{\theequation}
\edef\@savedtheequation{\the\@stequation}
\edef\oldtheequation{\theequation}
\def\theequation{\oldtheequation\alph{equation}}}
\def\endsubequations{\setcounter{equation}{\@savedequation}%
\@stequation=\expandafter{\@savedtheequation}%
\edef\theequation{\the\@stequation}\global\@ignoretrue

\noindent} \catcode`@=12
\begin{titlepage}

\title{{\bf  Running Spectral Index as a Probe of Physics at High Scales}} 
\vskip3in \author{ {\bf\sc G. Ballesteros}, {\bf\sc J.A. Casas} and
{\bf\sc J.R. Espinosa\footnote{\scriptsize\baselineskip=20pt E-mail 
addresses: 
{\tt guillermo.ballesteros@uam.es, alberto.casas@uam.es, 
jose.espinosa@cern.ch}}}
\hspace{3cm}\\
{\small IFT-UAM/CSIC, 28049 Madrid, Spain}.
}  \date{}  \maketitle  \def\baselinestretch{1.15}
\begin{abstract}
\noindent
The WMAP results on the scalar spectral index $n$ and its running with 
scale, though preliminary, open a very interesting window to physics at 
very high energies. We address the problem of finding inflaton 
potentials well motivated by particle physics which can accomodate WMAP 
data. We make a model independent analysis of a large class of models: 
those with flat tree-level potential lifted by 
radiative corrections, which cause the slow rolling of the inflaton and 
the running of $n$. This includes typical hybrid inflation models.
In the small-coupling regime the predictions 
for the size and running of $n$  are remarkably neat, 
{\it e.g.} $-dn/d\ln k=(n-1)^2\ll 1$, and $n$ does not cross $n=1$, 
contrary to WMAP indications. On the other hand, $n$ can run 
significantly if the couplings are stronger but at the price 
of having a small number of e-folds, $N_e$. We also examine the
effect of mass thresholds crossed during inflation.
Finally, we show that the presence of non-renormalizable
operators for the inflaton, suppressed by a mass scale above the
inflationary range, is able to give both $dn/d\ln k\sim {\cal 
O}(-0.05)$~and~$N_e\sim 50$.
\end{abstract}

\thispagestyle{empty}
\vspace*{0.1cm} \leftline{December 2005} \leftline{}

\vskip-18cm \rightline{IFT-UAM/CSIC-05-51} 
\rightline{hep-ph/0601134} \vskip3in

\end{titlepage}
\setcounter{footnote}{0} \setcounter{page}{1}
\newpage
\baselineskip=20pt

\noindent

\section{Introduction}

The measurement and analysis of the cosmic microwave background (CMB) by 
the WMAP collaboration \cite{WMAP,WMAP1,WMAP2} has provided results of 
unprecedented accuracy on the spectrum of the CMB anisotropies. The data 
are basically consistent with the main features of inflationary cosmology 
\cite{WMAP2}, implying important constraints for particular models (many 
of which can be discarded). One of the most intriguing WMAP results, 
though still uncertain, indicates that the spectral index of 
scalar perturbations, $n$, may exhibit a significant running with the 
wave-number, $k$.  More precisely, assuming a linear dependence of $n$ 
with $\ln k$, $n(k)=n(k_0)+[d n /d \ln k]\ln(k/k_0)$ ($k_0$ is some chosen 
``pivotal scale'', $k_0=0.002\ {\rm Mpc}^{-1}$), the analysis of 
ref.~\cite{WMAP2} gives $n(k_0)=1.13\pm 0.08$ and $d n /d \ln 
k = -0.055^{+0.028}_{-0.029}$ at 68\% C.L. 
(These values are obtained from WMAP data combined with different 
large-scale-structure data.) This means in particular that 
$n-1$ passes from positive to negative values in the range of $k$-values 
corresponding to these data ($10^{-4}\simlt k\ {\rm Mpc} \simlt 4$). It 
should be said, however, that a fit assuming constant spectral index is also 
consistent with the data, although somewhat worse, giving $n=0.99\pm 0.04$ 
(with WMAP data only \cite{WMAP}).

As a matter of fact, most inflationary models \cite{infl} do predict that 
$n$ runs 
with $k$, but the details of the ``observed'' running are very difficult 
to fit \cite{chung}. The last scales entering the observable universe 
(which we will denote $k_*\sim 10^{-4}\ {\rm Mpc}^{-1}$) 
should have crossed the horizon $\sim 
50-60$ e-folds before the end of inflation, at a time we will call 
$t_{*}$. The range of $k$ probed by the WMAP data corresponds to the first 
$\sim 6.5$ e-folds after $t_{*}$ (this range extends to $\sim 10.5$ 
e-folds if one adds large-scale-structure data).  During this ``initial'' 
period of inflation, $n-1$ should change sign (if the WMAP indication is 
confirmed), and it is not easy to cook up an inflationary potential 
implementing this feature (see \cite{chung,other} for an incomplete 
list of several analyses along 
these lines).  In the present paper we show, however, 
that a large and very interesting class of inflationary potentials,  
including those of many supersymmetric hybrid inflation models, are in 
principle able to accommodate this behavior quite naturally.

Let us recall a few basic results of inflationary cosmology in the 
slow-roll approximation. The slow-roll parameters are given by
\bea
\label{SlowRoll}
\epsilon={1\over 2}M_p^2 \left({V'\over V}\right)^2\ ,
\;\;\;\;\;\;
\eta=M_p^2 {V''\over V} \ ,\;\;\;\;\;\;
\xi=M_p^4 {V'V'''\over V^2} \ ,
\eea
where $8\pi M_p^2 = 1/G_{\rm Newton}$, $V$ is the scalar potential of the 
(slow-rolling) inflaton field, $\phi$, and the primes denote derivatives 
with respect to $\phi$. The slow-roll approximation requires $\epsilon, 
|\eta|, |\xi| \ll 1$, and the failure of this condition marks the end of 
inflation. The number of e-folds produced during the slow-rolling is 
given by
\bea
\label{Ne}
N_e = \int_{t}^{t_{\rm end}}H dt \simeq {1\over M_p^2}\int_{\phi_{\rm end}}
^\phi {V\over V'} d\phi = {1\over M_p}\int_{\phi_{\rm end}}
^\phi {1\over \sqrt{2\epsilon}} d\phi \ ,
\eea
where $H$ is the Hubble parameter [with $H^2=V/(3M_p^2)$], meaning that 
the scale parameter 
of the universe, $a$, grows as $a\rightarrow e^{N_e}a$. Successful 
inflation requires\footnote{It has been argued, however, that the 
required $50-60$ 
e-folds could be achieved through several steps of inflation, 
corresponding to different inflaton fields and/or inflaton potentials 
\cite{multinflation}. WMAP would only be sensitive to the first(s) of 
these steps.} 
$N_e\geq 50-60$. At first order in the slow-roll parameters 
\cite{makarov}, the power 
spectrum of scalar perturbations, $P_k$ (called $\Delta_{\cal 
R}(k)=(2.95\times 10^{-9})A(k)$ in ref.~\cite{WMAP2}), and  the spectral 
index, $n$, are given by 
\bea 
\label{Pk} 
P_k = {1\over 24\pi^2\epsilon}{V\over M_p^4}\ , 
\eea 
\bea 
\label{n} 
n\equiv 1 + {d \ln P_k \over d \ln k}\simeq 1+2\eta-6\epsilon\ . 
\eea 
Let us remind that, in the slow-roll approximation, the derivatives with 
respect to $\ln k$ can be related to field derivatives by using 
\bea 
\label{dlnk} 
{d \phi \over d \ln k}=-M_p^2{V'\over V}= -M_p\sqrt{2\epsilon}\ , 
\eea 
as is clear from (\ref{Ne}) and $a\propto 1/k$. It is then trivial
to obtain
\be
\label{dndlnk}
{d n\over d\ln k}\simeq -2\xi +16\epsilon\eta-24\epsilon^2\ .
\ee

The WMAP analysis \cite{WMAP1,WMAP2} gives 
\bea
\label{Pk0}
P_k=(2.95\times 10^{-9})\times(0.70^{+0.10}_{-0.11})\;\;\;\;  {\rm 
at}\;\;\; 
k=k_0\equiv 0.002\ {\rm Mpc}^{-1} \ ,
\eea
and $n$ as mentioned in the first paragraph:
\be
n(k_0)=1.13\pm 0.08\ ,\;\;\;\;   
{d n \over d \ln k} = -0.055^{+0.028}_{-0.029}\ .
\ee
Since usually $\epsilon\ll \eta$, then $n\simeq 1+2\eta$, and the 
change of sign in $n-1$ must
arise from a change of sign in $\eta$ (which implies in turn
that $V'$ must pass through an extremal point), something which, as 
mentioned
above, is non-trivial to implement (see ref.~\cite{chung} for a systematic 
analysis of this and related issues).

On the other hand, the evidence for a running $n$ comes from the first 
($l<5$) WMAP multipoles \cite{bridle}, which corresponds to the first 
$\sim 1.5$ e-folds of inflation (after $t_{*}$). Excluding those points, 
the fit is consistent with a constant index, $n\simeq 0.94\pm 0.05$, {\it i.e.}
the most significant running of $n$ seems to be associated to the first 
few e-folds and then $n$ remains quite stable. In any case, notice that
the errors associated to 
these analyses are big (actually refs.~\cite{WMAP2} and 
\cite{bridle} are only consistent within these errors).

Hence the results concerning the spectral index can be summarized by 
saying that they indicate
\bea
\label{n1}
{d n \over d \ln k}= {\cal O}(-0.05)\ ,
\eea
during the first few e-folds of inflation after $t_{*}$, and
\bea
\label{n2}
n\simeq 0.94\pm 0.05\ ,
\eea
during the following e-folds. Of course, if the previous
dramatic running is not confirmed by forthcoming analyses, $n$ could still
exhibit a milder running, compatible with (\ref{n2}) in the whole range
of observed scales.

In section~2 we study a large class of inflationary models defined by having
flat tree-level potential, with radiative corrections being responsible for the 
slow-roll of the inflaton and the running of $n$. The results depend on 
whether the couplings are weak (subsection~2.1) or strong (subsection~2.2) 
but 
generically one cannot get both a significant running of $n$ and a large 
$N_e$. 
We devote section~3 to a numerical analysis of this issue in the well 
motivated type of $D$-term inflation models.  Things improve in this 
respect if inflation is affected by some high energy threshold, 
possibility which 
is studied in section~4. Subsection~4.1 studies what happens if the 
inflaton crosses such high-energy threshold shortly after $t_*$ and 
subsection~4.2 analyzes the most promising case in which the threshold is 
above the inflationary range but affects inflation through 
non-renormalizable operators. In this last case we can easily accomodate 
both $dn/d\ln k\sim {\cal O}(-0.05)$ and $N_e\sim 50-60$. Finally, 
Appendix~A gives some technical details relevant to section~3 and 
subsection 4.1.

\section{Spectral index in a large class of inflation models}

Here we analyze a large class of inflationary models, namely those
with a flat inflaton scalar potential at tree-level. Then, the
flatness is spoiled by radiative corrections, which are responsible
for the slow-roll of the inflaton and, in turn, for the value and 
running of the spectral index. This class of models include many 
supersymmetric 
hybrid inflation models as a particularly interesting subclass (for other 
types 
see e.g. \cite{infl,treeflat}).
To be concrete, we will focus in the following on
hybrid inflation models, but all the results hold also
for generic tree-level-flat inflation models.

Hybrid inflation models \cite{hybrid} are among the most
popular and interesting ways to implement inflation. Their main
characteristic is that there are two (or more) scalar fields
involved in the inflationary process: the slow-roll field, $\phi$,
and the ``waterfall'' field(s), $H$. The scalar potential is such that, as 
long as $\phi$ is beyond a certain value, $\phi_c$, 
the minimum of the $H$-potential is at some ``false'' vacuum (typically
at $H=0$) and $\phi$ has an exactly flat potential at tree-level.
Below $\phi_c$ the minimum of the $H$-potential is at the true
vacuum. Once $H$ departs from the origin, the $\phi$ potential
is not flat anymore, inflation ends and $\phi$ goes to its true
minimum as well. The slow-roll of $\phi$ is driven by the radiative
corrections to the potential, which lift the tree-level flatness.
The main merit of such hybrid inflation is that the slow-roll parameters
$\epsilon$ and  $\eta$ can be easily small, as required, due to the 
loop-suppression factors and the fact that the scale of $\phi$ is not
directly related to the magnitude of $V$.

For our purposes, the main feature of hybrid inflation is that during
the inflationary period the tree-level potential, $V_0\equiv \rho = {\rm
constant}$, is
exactly flat for $\phi$, all its curvature arising from radiative
corrections. Generically, at one-loop
\bea 
\label{Vgeneric}
V(\phi) = \rho + \beta\ \ln{m(\phi)\over Q}\ ,
\eea
where $Q$ is the renormalization scale (which might have absorbed finite 
pieces). Note that $\rho$ depends 
implicitly on $Q$ through its renormalization group equation (RGE) and 
that the $Q$-invariance of the effective potential implies 
\be
\beta ={d\rho\over d \ln Q}\ ,
\ee
at one-loop. Finally, $m(\phi)$ is the most relevant
$\phi$-dependent mass [there could be several relevant (and different)
masses,  which is a complication we ignore for the moment].

The leading-log approximation (which amounts to summing the leading-log
contributions to all loops) is implemented in this context 
by simply taking $Q=m(\phi)$, so that
\be
\label{Vphi}
V(\phi) \simeq \rho(Q=m(\phi))  \ .
\ee
In general, one expects
$m^2(\phi) = M^2 + c^2 \phi^2$, where $M$ does not depend on $\phi$
and  $c$ is some coupling constant. We will ignore for the moment the
possible presence of the $M$ piece, as indeed happens in typical 
hybrid inflation models. From the equation $Q=m(\phi)$ we see that each 
value of $Q$ corresponds 
to a particular value of $\phi$ through\footnote{We take $c$ and 
$\phi$ as
positive, which is always possible through field redefinitions.} 
$\phi=Q/c(Q)$ (note that $c$ also depends on $Q$ according to 
its own
RGE).

The $\phi$-derivatives of the potential (\ref{Vphi}) and the 
corresponding slow-roll parameters can be easily calculated using
\bea
\label{derivphi}
{dQ\over d \phi}={Q\over \phi}\alpha= c \alpha\ ,
\eea
where
\bea
\label{alphabeta}
{1\over\alpha} = 1 - {\beta_c\over c}\;,\;\;\;\;\;{\rm 
and}\;\;\;\;\; 
\beta_c ={dc \over d \ln Q} \ .
\eea
In particular,
\bea
\label{epsiloneta}
V'&=& \alpha\ {\beta\over \phi} \ ,
\nonumber\\ 
V''&=& -\ \alpha\ {\beta\over
\phi^2} \left[1-\alpha{\dot\beta\over \beta} -\alpha^{2}\left(
{\dot{\beta_c}\over c}-{\beta_c^2\over c^2} \right) 
\right] 
\ ,
\eea
where the dots stand for derivatives with respect to $\ln Q$. 

Notice that from eq.~(\ref{dlnk}) we can relate the wave-number $k$ with 
the scale $Q$:
\be
\ln {k_0\over k}\simeq - {\rho\over \beta c^2M_p^2 }(Q_0^2-Q^2)
=- {3H^2\over \beta c^2}(Q_0^2-Q^2)\ . 
\ee
(This expression is exact in the small-coupling approximation.) Therefore, 
the spectral index at different $k$ scans physics at the corresponding 
(high-energy) scale. We analyze next this connection in detail.

\subsection{Small-coupling regime}

In the regime of very small coupling constants one has 
$\beta_c/c\ll 1$,
and thus $\alpha\simeq 1$, and we can also neglect all terms within the 
square 
brackets in (\ref{epsiloneta}), except the first one. Then
\bea
\label{epsiloneta2}
\epsilon\simeq {1\over 2}{M_p^2\over \phi^2}  \left({\beta\over
\rho}\right)^2\;,\;\;\;\;\;\; \eta\simeq -{M_p^2\over
\phi^2}{\beta\over \rho}\simeq  -2\left({\beta\over
\rho}\right)^{-1}\epsilon
\;,\;\;\;\;\;\; \xi\simeq 2\eta^2\ , 
\eea
so $\epsilon\ll\eta$, as usual, and thus $n-1\simeq 2\eta$. 

From the previous expressions it is clear that in this regime there exist 
very strong and model-independent relations between the observable parameters.
The running of $n$ with $\ln k$ is given by
\bea
\label{smallrunning}
{d n\over d \ln k}\simeq-2\xi\simeq
-4\eta^2 = -(n-1)^2\ ,
\eea
where we have used eqs.~(\ref{dndlnk}) and (\ref{epsiloneta2}). 
Consequently, ${d n/d \ln k}$
is negative, as suggested by observation, though its value
tends to be quite small. In fact the sign of $n-1$ cannot
change along the inflationary process, which would contradict the
WMAP indication of $n$ crossing the $n=1$ value.
The sign of ${d n/d \ln k}$
and the correlation of its absolute value with that of $n-1$
are quite model-independent: they apply whenever the field dependence of the 
inflaton potential is well described simply by a $\ln\phi$. If 
the value of $\beta$ 
changes along the inflationary course, e.g. if some threshold of extra
particles is crossed in the way, this correlation still holds except in 
the neighborhood of the thresholds.
The signs of  the first derivative of the potential, $V'\simeq 
\beta/\phi$, and $\eta$ (and thus $n-1$) are also correlated; namely 
\bea
\label{signs}
{\rm
sign}(n-1)=-{\rm sign}(V')=-{\rm sign}(\beta)\ .
\eea
Notice that a change in the sign of $n-1$ can only occur if $V'$  
vanishes, indicating the presence of a local minimum.
Usually $\beta$ (and thus $V'$) is positive and therefore we naturally 
expect $n<1$. 
Then $\phi$ rolls towards the origin, where the true minimum of the 
$\phi-$potential is in typical hybrid inflation 
models \cite{hybrid}. Inflation ends either because the flatness condition
gets violated below some critical value $\phi_c$ (as mentioned above) or
because $\epsilon$ or (much more often) $|\eta|$ become ${\cal O}(1)$.
In this respect, 
notice that $\epsilon$ and $|\eta|$ increase as $\phi$ rolls to smaller
values. On the contrary, if $\beta$ (and thus $V'$) is negative, 
then $n>1$, $\phi$
runs away and the slow roll parameters $\epsilon$ and $\eta$ get
smaller as time goes by. Inflation would not stop unless
$\phi$ reaches some critical value where the tree-level flatness ends and
all the fields go to the true minimum. This is not the most common 
situation, but it cannot be excluded (see {\it e.g.} \cite{Natural,IR,BCH}).

In the usual case, with inflation ending because $\eta$ gets 
${\cal O}(1)$ (which according to the previous discussion requires 
$\eta<0$) and in the absence of thresholds of new physics during 
the inflationary 
process\footnote{A detailed study with thresholds is performed in 
subsection~4.1.} the number of e-folds, $N_e$, since $t_*$ until the end 
of inflation can be easily computed plugging (\ref{epsiloneta2}) in 
eq.~(\ref{Ne}) and is simply
\bea
\label{Nstar}
N_e\simeq -{1\over 2}
\left[{1\over\eta(\phi_*)}-{1\over\eta(\phi_{end})}\right]
\simeq  -{1\over 2\eta(\phi_*)}\ .
\eea
If there are no other steps of inflation (see footnote 1),
then $N_e=50-60$; otherwise it could be smaller.
Then the value of the spectral index at the beginning of inflation is 
given by
\bea
\label{nstar}
n_*\equiv n(k_*)\simeq 1+2\eta(\phi_*)\simeq 1 - {1\over N_e}\ ,
\eea
which means $n_*\leq 0.983$ (the equality holds for $N_e = 60$, 
i.e. if there are no additional steps of inflation). For a given
value of $\beta/\rho$ this also fixes the initial value $\phi_*$ 
through eq.~(\ref{epsiloneta2}). In addition, we can use
(\ref{nstar}) to write eq.~(\ref{smallrunning}) in an integrated form as
\bea
\label{smallrunning2}
n =1 -{1\over N_e - \ln (k/ k_*)} 
\ .  
\eea
For later use, it is convenient to rewrite this in terms of 
the initial value of the spectral index, $n_*$ given by eq.~(\ref{nstar}), 
as
\bea
\label{nintegrated}
n = n_* + {1\over N_e} - {1\over N_e - \ln (k/ k_*)}\ .
\eea

If inflation ends by the breaking of the flatness condition, 
eqs.~(\ref{nstar}--\ref{nintegrated}) hold, 
replacing $N_e\rightarrow -1/[2\eta(\phi_{*})]$.
This also holds for the $\beta<0$ (and thus $n>1$) case.
Finally, note that there is no problem in reproducing the observed
scalar power spectrum (\ref{Pk0}): for a given
value of $\beta/\rho$, this can be achieved with an appropriate
value of $V\equiv\rho$, according to
eq.~(\ref{Pk}).

In summary, in the small coupling regime the predictions of tree-level-flat
inflation models (including many hybrid inflation models) on 
the size and running 
of the spectral index are quite
neat and model-independent. 
Forthcoming WMAP analyses could either confirm or falsify them
(e.g. if the crossing of $n$ through $n=1$ is verified).

\subsection{Not-so-small-coupling regime}

If the $\beta$-functions are positive (as usual, see the above discussion),
couplings grow in the ultraviolet and there will be a scale where the 
second and third terms within the square brackets in (\ref{epsiloneta}) 
compete with the first one. Since they naturally have the opposite sign, 
one can expect that at sufficiently high 
scales (which means initial stages of inflation) the sign of $\eta$, and
thus $n-1$, may get positive, which therefore could account for the
preliminary indication of WMAP.
Notice in particular that, since eq.~(\ref{epsiloneta2}) for $\eta$ is not 
 a valid
approximation anymore, the previous correlation (\ref{signs}) between 
the signs of $V'$
and $\eta$ does not hold, which allows $\eta$ to pass from positive
to negative values without the presence of an extremal point of the 
potential $V$. As a matter of fact, we find this behaviour absolutely
natural, and even unavoidable in this regime. However, to reproduce
the WMAP preliminary indication on the running spectral index {\em and} 
the other physical requirements at the same time (in particular, a 
sufficient number of e-folds) is not a trivial matter. This will be 
analyzed for typical examples in sect.~3.

The running of the spectral index can be 
computed, taking into account that, as long as $\beta/\rho\ll 1$,
 still $\epsilon\ll |\eta|$
(except within the close neighborhood of the point where
$\eta$ vanishes), so $n\simeq 1+2\eta$, and
\bea
\label{largerunning}
{d n\over d\ln k}\simeq 2{d \phi\over d\ln k}{d \eta\over d\phi}=
-2M_p\sqrt{2\epsilon}{d \eta\over d\phi}
\ ,
\eea
where $d \eta/ d\phi$ can be straightforwardly evaluated using the whole 
expression for $\eta$ from (\ref{epsiloneta}), taking into account the 
implicit dependence on $\phi$ through the one on $Q$ with the help of 
eq.~(\ref{derivphi}). Similarly, the number of e-folds is computed using 
eqs.~(\ref{Ne}, \ref{epsiloneta}). In the next section we will show 
explicit expressions for these quantities in particular examples.

\section{An explicit example}

In the previous section we claimed that, within the hybrid inflation framework,
it is natural to expect a running spectral index, even crossing the
$n=1$ value, if the scale of the initial stages of inflation
is high enough. Now we will explicitly show that this is the case by
working out a popular example of hybrid inflation, namely the 
first example of hybrid D-term inflation, proposed by Bin\'etruy and 
Dvali \cite{BD}.

The model is  globally supersymmetric with canonical K\"ahler
potential, and with superpotential\footnote{We follow here the formulation
with a tadpole for $\Phi$ in the superpotential as in ref.~\cite{BCH}, 
where this model was analyzed in connection with the issue of 
decoupling.}
\bea
\label{W}
W=\Phi(\lambda H_+H_--\mu^2)\ ,
\eea
where $\lambda$ and $\mu^2$ are real constants and 
$\Phi, H_+, H_-$ are chiral superfields.
$H_\pm$ have charges $\pm 1$ with respect to a $U(1)$ gauge group
with gauge coupling $g$ and a Fayet-Iliopoulos term $\xi_D$. 
The associated tree-level scalar potential is given by $V_0 = 
V_F +V_D$,
with
\bea
\label{VFVD}
V_F&=&\left|\lambda H_+H_--\mu^2\right|^2 
+ \lambda^2\left(\left|H_-\right|^2 
+ \left|H_+\right|^2\right)|\Phi|^2\ ,\nonumber\\
V_D&=&{g^2\over 2}\left(\left|H_+\right|^2 
- \left|H_-\right|^2
+ \xi_D\right)^2\ ,
\eea
(the fields in these expressions are the scalar components of the 
chiral superfields).
The global minimum of the potential
is supersymmetric
(i.e. $V_F=0$, $V_D=0$)
and occurs at 
$\Phi=0$, $|H_\pm|^2=(\mp\xi_D +\sqrt{\xi_D^2+ 4\mu^4/\lambda^2})/2$.
But for large enough $|\Phi|$ 
\bea
\label{phic}
|\Phi|^2> |\Phi_c|^2=
{1\over \lambda^{2}}\Delta\ ,
\;\;\;\;\Delta\equiv\sqrt{g^4\xi_D^2+\lambda^2 \mu^4}\ ,
\eea
the potential has a minimum at $H_\pm=0$ and
is flat in $\Phi$. This flat direction is two dimensional because $\Phi$ 
is complex, $\Phi=(\phi_R+i\phi_I)/\sqrt{2}$. In the following, we examine 
the potential along the real part $\phi\equiv\phi_R$. All our results 
would be the same along any other radial direction. It is in this 
flat region where hybrid inflation takes 
place. The tree-level potential during this period is simply
\bea
\label{rhoBD}
V_0\equiv \rho=\mu^4 + {1\over 2}g^2\xi_D^2\ .
\eea
The one-loop correction is given by
\bea
\label{V1BD}
\Delta V_1 = {1\over 32 \pi^2}\sum_{i=\pm}\left[
\tilde m_i^4\ln{\tilde m_i^2\over Q^2}\ -\  m_i^4\ln{m_i^2\over Q^2}
\right]\ ,
\eea
where $m_i^2$ ($\tilde m_i^2$) are the $\phi$--dependent square masses
of the fermionic (scalar) components of the $H_\pm$ superfields:
\bea
\label{BDmasses}
m_\pm^2= m^2(\phi) \equiv {1\over 2}\lambda^2 \phi^2\ ,
\;\;\;\;\;\;\;
\tilde m_\pm^2=m^2(\phi)\pm\Delta\ .
\label{BDmasses2} 
\eea
In practice, $\tilde m_\pm^2$ is dominated by the $m^2(\phi)$ 
contribution. To see this, notice that the condition $m^2(\phi)=\Delta$ 
(i.e. $\tilde m_-^2=0$) precisely corresponds to the point $\phi=\phi_c$,
below which the flatness of the potential breaks down and inflation cannot
continue.
Since the number of e-folds goes as $\sim \phi^2$, most of the e-folds
have been produced at larger (usually much larger) values
of $\phi^2$, where $m^2(\phi)\gg\Delta$. 
This is particularly true in the region of interest for us, i.e.
the initial stages of inflation. 
Furthermore, as we will see below, inflation ordinarily
ends before $\phi^2=\phi_c^2$, namely when
$\eta$ becomes
${\cal O}(1)$, which normally occurs for much larger $\phi^2$.
In consequence, $\Delta V_1$ can be written as
\bea
\label{V1BD2}
\Delta V_1 = {1\over 8 \pi^2}\Delta^2
\ln{m(\phi)\over Q}\;+\; {\cal O}\left({\Delta^2\over m^4(\phi)}\right)
\; , 
\eea
so $V=V_0+\Delta V_1$ is of the generic form shown in 
eq.~(\ref{Vgeneric}).

The various $\beta$-functions, defined as derivatives with respect to $\ln Q$, 
are given by
\bea
\label{betaBD}
\beta&=&{1\over 8\pi^2}\Delta^2\ ,\\
\label{betae}
\beta_g&=&{1\over 8\pi^2} g^3 \ ,\\
\beta_{\xi_D}&=& 0\ ,\\
\label{betag}
\beta_\lambda&=&{1\over 16\pi^2}\lambda(3\lambda^2-4g^2)\ ,\\
\label{betagv}
\beta_{\mu^2}&=&{1\over 16\pi^2} \lambda^2 \mu^2\ .
\eea
In the leading-log approximation the radiatively-corrected potential
is simply given by $\rho$ evaluated at the scale 
$Q=m(\phi)=\lambda\phi/\sqrt{2}$; 
in the notation of sect.~2, 
\be
c\equiv \lambda/\sqrt{2}\ .
\ee

Concerning the end of inflation, this occurs at $\phi_{\rm end} = 
{\rm Max}\{\phi_c, \phi_\eta\}$, where $\phi_c$
marks the end of tree-level flatness and is given in eq.~(\ref{phic}), 
and $\phi_\eta$ corresponds to the point where $\eta={\cal O}(1)$.
Using eq.~(\ref{epsiloneta2}) for an estimate,
\be
\label{phicp}
\phi_\eta^2\simeq  M_p^2\ {\beta\over \rho}\ .
\ee
It is easy to see that $\phi_c<\phi_\eta$ for a wide range of $\{g, 
\lambda\}$ so that, normally, inflation ends
when $\eta={\cal O}(1)$. E.g. for $g\sim \lambda$ 
and using eqs.~(\ref{Pk}) and (\ref{Pk0}), taking into account that 
$\phi(k_0)$ in (\ref{Pk}) and  (\ref{Pk0}) should be
at least equal to $\phi_\eta$, one finds that this occurs 
for $g, \lambda\simgt 9 \times 10^{-2}$. More realistically,
when $\phi(k_0)\sim M_p$, this occurs for $g, \lambda\simgt 10^{-3}$.

Now, a successful inflationary process, consistent with the WMAP constraints,
should satisfy the following requirements

\begin{description}

\item[{\em i)}] At least 50--60 e-folds of inflation.

\item[{\em ii)}] A correct power spectrum of the scalar perturbations at
$k_0=0.002\ {\rm Mpc}^{-1}$, as given by eqs.~(\ref{Pk}) and (\ref{Pk0}).

\item[{\em iii)}] Running spectral index with $n-1$ passing from 
positive to negative
during the first few e-folds of inflation (after $t_{*}$), with ${d n / d 
\ln k}= {\cal O}(-0.05)$.

\end{description}
As discussed in the Introduction, the experimental status of requirement 
{\em iii)} is still
uncertain, but we will examine here the possibilities of fulfilling it
in this context. Likewise, condition  {\em i)} can be relaxed if 
there are several steps of inflation, as discussed in footnote 1.
In the latter case 7--10 e-folds 
(i.e. the range covered by WMAP with the optional
complement of additional observations) might be acceptable.

If $g$ and $\lambda$ are small, $\beta$ remains essentially constant (and 
positive) during inflation. Then the results of sect.~2 apply, in 
particular $n<1$, with a slight running $d n /d \ln k<0$, as given by 
eqs.~(\ref{smallrunning}), (\ref{nstar}) and (\ref{smallrunning2}). 
According to that discussion, it is not possible in this regime to fulfill 
condition {\em iii)}, i.e. a running spectral index crossing through 
$n=1$. 

In consequence we will focus for the rest of this section in
the case of more sizeable coupling constants. Let us first
verify that the slow roll conditions are fulfilled.
In this regime $\epsilon$ and $\eta$ are deduced from
eqs.~(\ref{epsiloneta}).
Taking into account that $\alpha={\cal O}(1)$ and that $\dot{\beta}$
is at most ${\cal O}(\beta)$, it is clear that $\eta = {\cal O}(
M_p^2\beta/\rho\phi^2)$ and $\epsilon =
{\cal O}(\eta\beta/\rho)$. Therefore $\epsilon, |\eta|\ll 1$
as long as 
\bea
\label{slow}
{\beta\over \rho}\ll {\phi^2\over M_p^2}\ ,
\eea
which naturally leads to $\beta\ll\rho$. A simple way of guaranteeing
this is to take
\bea
\label{condition}
{g^2\over \lambda^2}\ll {\mu^4\over g^2\xi_D^2}\ll {1\over 2}\ ,
\eea
{\it i.e.} $\beta$ is dominated by the Yukawa contribution:
\bea
\label{dom1}
\beta\simeq {\lambda^2\mu^4\over 8\pi^2} \ ,
\eea
{\em and} $\rho$ is dominated by the D-term (Fayet-Iliopoulos) 
contribution:
\bea
\label{dom2}
\rho\simeq {1\over 2}g^2\xi_D^2\ .
\eea
Imposing (\ref{condition}) is by no means the only way of satisfying the 
slow-roll conditions [e.g. reversing the inequalities in 
(\ref{condition}) would also work], but it is particularly simple and 
useful for a clear analytic discussion\footnote{Actually, once global
supersymmetry is promoted to supergravity, requiring $\mu^4 \ll 
g^2\xi_D^2/2$ [i.e. the second inequality of (\ref{condition}), and thus
eq.~(\ref{dom2})] is mandatory for almost any form of the 
K\"ahler potential, $K$.
The reason is the well-known $\eta-$problem, i.e. the appearance of 
an ${\cal O}(H^2)$ mass term for the inflaton (and thus $\eta\sim 1$) 
if $V_F\sim V$. However, it is remarkable that for the superpotential
of this model, eq.~(\ref{W}), and minimal  K\"ahler potential, 
$K=|\Phi|^2$,
no such mass term is generated due to an amusing cancellation of 
contributions.}.

Using this dominance of the Yukawa coupling, we can evaluate $\epsilon$ 
and $\eta$ from eqs.~(\ref{epsiloneta})
\bea
\label{alphaepsiloneta2}
\epsilon&=& {1\over 2}\alpha^{2}
{M_p^2\over \phi^2}
\left({\beta\over \rho}\right)^2 \ ,
\nonumber\\ 
\eta&=&\alpha  {M_p^2\over
\phi^2}
{\beta\over\rho}
\left[-1 + {\lambda^2\over 2\pi^2}
\left(\alpha + {9\over 32}{\lambda^2\over 2\pi^2}\alpha^{2}
\right)\right] \ ,
\eea
with
\bea
\label{alphabeta2}
{1\over\alpha} = 1 - {\beta_\lambda\over \lambda}= 1-{3\lambda^2\over 
16\pi^2}\ .
\eea

As announced in the previous section, there is a scale, say $Q_0$, 
at which $\eta\simeq 0$ and $n-1$ passes from positive to negative.
From (\ref{alphaepsiloneta2}), the corresponding value of $\lambda$,
$\lambda_0\equiv\lambda(Q_0)$, turns out to be independent of the other 
parameters of the model, namely
\bea
\label{l0}
{\lambda_0^2\over (4\pi)^2}={1\over 15}(7-\sqrt{34})\simeq 0.078\ .
\eea
The coupling $\lambda_0$ is quite sizeable ($\lambda_0\simeq 3.51$), but 
well within the perturbative range, 
$\lambda_0^2/(4\pi)^2\ll 1$. It is perfectly possible that $Q_0$
is crossed during the first few e-folds after $t_{*}$ 
(recall $Q= \lambda\phi/\sqrt{2}$), so that this 
hybrid inflation model (and in general any tree-level-flat inflation 
model) is able to implement the crossing of the spectral index through
$n=1$ at small $k$.
However, to reproduce the suggested value of $d n /d \ln k$
{\em and} a sufficient number of e-folds at the same time 
[i.e. conditions {\em i), iii)} above] is not that easy, as we show 
next.

From 
eq.~(\ref{Ne}), the number of 
e-folds between $Q_0$ and the end of inflation can be written as
\bea
\label{Ne2}
N_e^{0} &=&  {1\over  M_p^2}\int_{Q^2_{\rm end}}
^{Q^2_0} {1\over \alpha^2\lambda^2}{\rho \over \beta} dQ^2
\nonumber\\
&&\hspace{1cm}
\nonumber\\
&\simeq&
-{64\pi^4\over 3}
\left[{Q_0^2\over M_p^2} {g^2\xi_D^2\over \mu_0^4}\right]
\lambda_0^{2/3}{\rm exp}\left[{16\pi^2\over 3\lambda_0^2}\right]
\int_{1/\lambda^{2}_{\rm end}}^{1/\lambda^{2}_0}  
x^{7/3} 
\left(1-{3\over 16\pi^2x}\right)^2
{\rm exp}\left[-{16\pi^2x\over 
3}\right]
dx\nonumber\\
&&\hspace{1cm}
\nonumber\\
&\simeq& 0.303\ \left[{Q_0^2\over M_p^2}\ 
{g^2\xi_D^2\over \mu^4_0}\right]
=1.818\ \left[{Q_0^2H^2\over \mu^4_0}\right]
\ ,
\eea
where $\mu_0=\mu(Q_0)$. In the first line  we have performed a change of 
variable
$\phi\rightarrow Q^2$ with respect to the expression (\ref{Ne}). In the 
second line we have changed variables once more, 
$Q^2\rightarrow x=1/\lambda^{2}$,
neglecting the $\propto g^2$ contributions to the RGEs of
$\lambda$ and $g^2$ as well as 
the $\mu^4$ contribution
to $\rho$, according to assumption (\ref{dom2}). Finally, in the
last line we have taken into account that the contribution
to the integral coming from small values of $\lambda$ is negligible, so we
have replaced $\lambda_{\rm end}\rightarrow 0$, $\lambda_0$ 
by the value quoted
in (\ref{l0}), and evaluated the integral numerically. The final
result turns out to be a very good approximation, provided 
eq.~(\ref{condition}) is fulfilled.
Notice that, since the point $Q_0$ (i.e. where $n=1$) should be
crossed after the first few e-folds of inflation, $N^0_e$ should
be ${\cal O}(50)$ [unless there are subsequent episodes of inflation, 
in which case it could be as small as ${\cal O}(1)$].

On the other hand, the running $dn/d \ln k$ can be straightforwardly 
evaluated using (\ref{dndlnk}). It gets a particularly simple form when 
evaluated at the scale $Q_0$, where $\eta=0$, $n\simeq 1$:
\bea
\label{detalng0}
\left.{dn\over d \ln k}\right|_{Q_0}&=&
-{3\over 4}\lambda_0^4\alpha_0^5(3\alpha_0-1)
\left({\lambda_0^2\over 4\pi^2}\right)^4
\left[{Q_0^2\over M_p^2}\ {g^2\xi_D^2\over 
\mu^4_0}\right]^{-2}\nonumber\\
&&\nonumber\\
&\simeq& -11.847 \left[{Q_0^2\over M_p^2}\ {g^2\xi_D^2\over 
\mu^4_0}\right]^{-2}=-0.33\left[{Q_0^2H^2\over \mu^4_0}\right]^{-2}
\ ,
\eea
where we have used (\ref{l0}) for the numerical estimate. Clearly, 
$N_e^0$ and $\left.{dn/ d \ln k}\right|_{Q_0}$ depend
on the same combination of input parameters (the one in square brackets);
and the product
$(N_e^0)^2\times (\left.{dn/ d \ln k}\right|_{Q_0})$ turns out to be 
independent of them
\be
\label{rule}
-(N_e^0)^2\times \left.{dn \over d \ln k}\right|_{Q_0}\simeq 1.1\ .
\ee
This constraint makes it impossible to fulfill 
$\left.{dn/ d \ln k}\right|_{Q_0}={\cal O}(-0.05)$ and
$N_e^0\sim {\cal O}(50)$ simultaneously.
However $N_e^0\sim {\cal O}(5)$
is perfectly possible, which means that the model can account
for the indication of WMAP, provided there exist subsequent episodes
of inflation.

\begin{figure}[t]
\vspace{1.cm} \centerline{
\psfig{figure=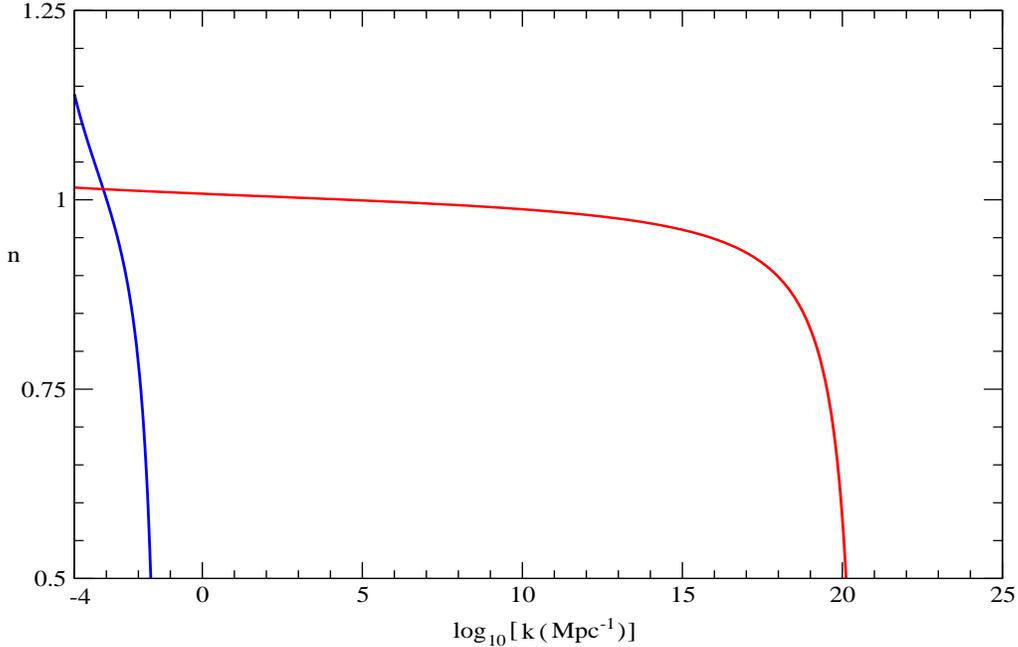,angle=-90,height=8cm,width=8cm,bbllx=4.cm,%
bblly=7.cm,bburx=20.cm,bbury=21.cm}}
\caption{\footnotesize 
Scalar spectral index as a function of ${\rm log}_{10}[k({\rm Mpc}^{-1})]$
for the inflaton potential of eqs.~(\ref{rhoBD}) and (\ref{V1BD}) for two 
different choices of parameters: one to get right $N_e$ and the other to 
get right $dn/d\ln k$.} 
\label{fig:nLOG}
\end{figure}

The previous behaviour is illustrated by Fig.~\ref{fig:nLOG} which gives 
the spectral index $n$ as a function of the scale wave-number $k$. 
The flat red curve  
[corresponding to $g_*=0.02$, $\lambda_*=4$, $\mu_*^2=(3.6\times 10^{-4} 
M_p)^2$, $\xi_D=(1.2\times 10^{-2} M_p)^2$, $\phi_*=0.28 M_p$] is able to 
accomodate a large $N_e\simeq 57$ but has very small running. 
The steep blue curve instead [with the same parameters except for
$\mu_*^2=(1.9\times 10^{-3} M_p)^2$ and $\xi_D=(3.6\times 10^{-2} M_p)^2$] 
has a sizeable $dn/d\ln k\simeq -0.075$ but fails to give enough e-folds: 
$N_e\simeq 7$.

Interestingly, there is no problem in reproducing the observed
scalar power spectrum (\ref{Pk0}). Notice that $P_k$ in eq.~(\ref{Pk})
is now
\bea
\label{Pk00}
\left.P_k\right|_{Q_0}&=& {4\pi^2\over 3}{1\over \alpha_0^2\lambda_0^6}
{(g^2\xi_D^2)^2\over M_p^4 \mu^4_0}
\left[{Q_0^2\over M_p^2}\ {g^2\xi_D^2\over \mu^4_0}\right]
\nonumber\\
&&\nonumber\\
&\simeq & 4.14\times 10^{-4} 
{(g^2\xi_D^2)^2\over M_p^4 \mu^4_0}
\left[{Q_0^2\over M_p^2}\ {g^2\xi_D^2\over \mu^4_0}\right]
=0.089{H^4\over \mu^4_0}
\left[{Q_0^2H^2\over \mu^4_0}\right]
\ .
\eea
Clearly, for any given value of the quantity  in the
square brackets, it is possible to vary the parameters
so that the prefactor of (\ref{Pk00}) changes in order 
to fit the observed $P_k$. Incidentally, the pivotal scale 
in the analysis of ref.~\cite{WMAP2}, $k_0 = 0.002\ {\rm
Mpc}^{-1}$ (which corresponds to $\sim 3$ e-folds) 
approximately coincides with the stage
at which $n\simeq 1$ (i.e. when $Q=Q_0$). This is
a fortunate circumstance to perform rough estimates.

In summary, the simple hybrid inflation model of ref.~\cite{BD}, 
in the small coupling regime, produces the general results shown in 
subsect 2.1, with
very clean predictions, in particular $n<1$ with a mild
${dn/ d \ln k}<0$ running. In the regime of sizeable couplings, 
the model can
account for the scalar power spectrum at the pivotal scale and
a running spectral index crossing $n=1$ at a suitable
scale with the suggested slope,
${dn/ d \ln k}={\cal O}(-0.05)$.
However, the number of e-folds produced is rather small, 
so this possibility is not viable unless there are
subsequent episodes of inflation.

The previous analysis can be repeated for an arbitrary number of pairs of 
$H_\pm$ fields, using the formulae given in Appendix~A. The results do not 
change much, except for the fact that the value of $\lambda_0$ scales
as $1/\sqrt{N}$. Also, the right hand side of eq.~(\ref{rule}) depends 
slightly on the value of $N$, being always $\simlt 1.1$.

Of course, one cannot exclude that other tree-level-flat
models are able to reproduce such running spectral index without the 
above limitation for the number of e-folds. To achieve this
one needs sizeable $\eta$ at high scale, so that 
the spectral index runs apreciably in the the first stages of inflation,
{\em and} small $\epsilon$ at lower scales,
so that a sufficient number of e-folds is produced. This implies
that $\beta$ (and thus the relevant coupling constants) 
must evolve from sizeable values at high scale
to substantially smaller values at lower scales. The faster such 
evolution,
the easier to accomplish both requirements. Since this evolution
is a result of the RG running, one can  imagine
scenarios where it is speeded up. For example, if some 
threshold of new physics is crossed in the inflationary
course, the $\beta$-functions
may suffer quick changes, as desired. However the analysis
has to be done carefully to eliminate spurious effects, as
we shall see shortly. On the other hand, the existence of new physics can 
have other (even more) interesting implications.

In the next section we analyze carefully the various effects of 
thresholds of new physics in the inflationary process.

\section{Thresholds of new physics}

The Lagrangian responsible for the inflationary process can well be
an effective theory of a more fundamental theory at a higher scale.
Actually, since during inflation the characteristic scale is
changing (following the value of the inflaton field), it may even
happen that some threshold of new physics, say $\Lambda$, is crossed in 
that process. In either case one can expect effects from the new physics in
the details of inflation.

For the kind of inflationary models  we are considering in this paper,
namely tree-level-flat models, there can be two main types of new
physics: 1) new physics that  does not spoil the tree-level flatness
of the potential and 2) new physics that does spoil it. In case 1), 
the new physics can only affect the form of the initial Lagrangian through
radiative corrections.  If $\Lambda$ is above the scales of the whole
inflationary process, these effects are suppressed by the ratio of the
scales, $\sim (\phi/\Lambda)^n$, besides the typical loop suppression
factors. Thus we expect them to be much less important than the
radiative effects calculated  within the effective theory considered
in previous sections.  On the other hand, if $\Lambda$ is crossed in the 
inflationary course, the effective potential and the various RGEs involved
will get modified, which can have interesting implications, as mentioned
at the end of sect.~3. In contrast,
for case 2) the new physics can show up, even below
$\Lambda$, through non-renormalizable operators in the tree-level 
potential,
suppressed by  $\sim (\phi/\Lambda)^n$. Actually, in case 2) it is not
realistic to suppose that $\Lambda$ is crossed during inflation, since
above $\Lambda$ the new physics will usually spoil the slow-roll conditions.
Next we will explore possibilities 1) and 2) in turn.

\subsection{Thresholds crossed in the inflationary process}

Let us suppose that the inflaton potential is flat at tree-level
and the 1-loop radiative correction, $\Delta V_1$, contains
contributions from states with different masses. This generalizes the 
situation discussed in sect.~2.
To be concrete, consider two types of particles with masses
$m(\phi)$ and $M(\phi)$ satisfying $m(\phi)\ll M(\phi)$. The 1-loop 
effective
potential reads
\bea 
\label{Vfull1}
V(\phi) = \rho(Q) + \beta_l\ln{m(\phi)\over Q} + 
(\beta_h-\beta_l)\ln{M(\phi)\over Q} \ .
\eea
The tree-level potential, $\rho$, 
depends implicitly on $Q$, as indicated, with RGE
corresponding to the full theory, $d \rho/d\ln Q=\beta_h$.
Notice that $\beta_l$ is the corresponding $\beta-$function
when only the light particles are present. Generically, 
$m^2(\phi)=m^2+c^2\phi^2$, $M^2(\phi)=M^2+C^2\phi^2$, where 
$m$ and $M$ do not depend on $\phi$, and 
$c$ and $C$ are coupling constants.

As mentioned above, we focus on the interesting situation
in which the scale of inflation
crosses the threshold $M$ where the heavier particles decouple. 
In that case one needs a reliable prescription to evaluate
$V(\phi)$ above and below the threshold. In particular one should
take care of potentially large logarithms in the potential (\ref{Vfull1}) 
and higher order corrections to it.
In the case of just one type of mass, say $m(\phi)$, this can be  achieved
by choosing $Q=m(\phi)$, so that the unique log in the radiative 
correction cancels and $V(\phi)=\rho[Q(\phi)]$. This corresponds to 
summing up the leading-log
contributions in $V$ to all loops. This simple procedure
is not possible here because of the two different logs in (\ref{Vfull1}). 
Then, a convenient approach is to perform the 
integration of the RGEs in an effective theory framework.
To this end, we consider 
two different regions for the potential. In the high energy region, when 
$\{Q,m(\phi)\}\geq M$, the potential, $V_{high}$, 
is as written in (\ref{Vfull1}), with 
RGEs 
including the virtual effects of the heavy states. However, in the low 
energy region, defined by $\{Q,m(\phi)\}\leq M$, the potential,  
$V_{low}$,
is defined by dropping the contribution from the 
heavy multiplets to the potential and to all RGEs. For this procedure to 
be consistent
one has to match the high- and low-energy potentials, i.e.
one has to add a piece $\delta_{th}V_{low}(\phi)$ to $V_{low}$
to guarantee $V_{low}=V_{high}$ at $Q=M$. In this way one gets
\be
\label{Vlow}
V_{low}(\phi)=\rho(Q) + \beta_l\ln{m(\phi)\over Q}\ +\ 
\delta_{th}V_{low}(\phi) ,
\ee
where $\rho$ now runs with the low energy RGE, i.e. with $\beta_l$, and
\be
\delta_{th}V_{low}(\phi)=
\left.(\beta_h-\beta_l)\right|_M\ln{M(\phi)\over M}\ ,
\ee
where the subindex $M$ means that the quantity within the parenthesis is 
evaluated at $Q=M$. The additional piece $\delta_{th}V_{low}(\phi)$ can be 
expanded in powers of $\phi/M$ if desired: one would get $\phi$ operators 
suppressed by inverse powers of $M$, which is the reason one can discard 
such high energy remnants when interested in low energy physics. However 
we keep here the whole expression as we are also interested in the 
potential for values of $\phi$ in the neighbourhood of the threshold. 
Actually a correct treatment of that region is crucial for a reliable 
analysis. Notice for instance that since $\epsilon\sim 1/\phi$ most 
of 
the e-folds in the energy region below $Q=M$ are produced precisely near 
that 
threshold. The inclusion of the $\delta_{th}V_{low}(\phi)$ contribution 
usually increases $\epsilon$, therefore decreasing (importantly) the 
number of low-energy e-folds, and should not be neglected.

The next step is to make a judicious choice of the renormalization scale $Q$.
The simplest option is to take $Q(\phi)=m(\phi)$, which works well in 
both energy regimes (meaning that the explicit logarithms are never large). 
After doing that we get the following expressions for the potential in the two 
regions:
\bea
V_{high}(\phi)&=&\rho[Q(\phi)]+
(\beta_h-\beta_l)\ln{M(\phi)\over m(\phi)}\ ,
\nonumber\\
V_{low}(\phi)&=&\rho[Q(\phi)]+\left.(\beta_h-\beta_l)\right|_M
\ln{M(\phi)\over M}\ .
\label{VHVL}
\eea

As a matter of fact, the previous approximation may not be accurate enough
in some cases, {\it e.g.} when the coupling constants involved are not 
so small. In particular, although $V$ in (\ref{VHVL}) is continuous 
across the threshold
by construction, the derivatives $V'$, $V''$ (and thus $\epsilon$ and $\eta$)
are not. This is an spurious effect which is conveniently smeared out when
the approximation is improved. This can be done by including higher-loop
corrections to the effective potential. Indeed, without
any further calculations, one can obtain the leading-log corrections
at arbitrary loop order by using the $Q$-invariance of the potential.
In particular, at 2-loop leading-log (2LL) the potential reads
\bea 
V^{2LL}(\phi)\ &=& \rho(Q) + \beta_l\ln{m(\phi)\over Q} + 
(\beta_h-\beta_l)\ln{M(\phi)\over Q} 
\nonumber\\
&+& {1\over 2}\dot\beta_l\left[\ln{m(\phi)\over Q}\right]^2
+ {1\over 2}\left(
\cbeta_h-2\cbeta_l+\dot\beta_l\right)
\left[\ln{M(\phi)\over Q}\right]^2
\nonumber\\
&+&\left(\cbeta_l-\dot\beta_l\right) \ln{m(\phi)\over 
Q}\ln{M(\phi)\over Q}\ ,
\label{Vfull2}
\eea
where the circle (dot) indicates $d/d\ln Q$ in the high (low) 
energy region. 
Redoing the decoupling and matching procedure at 
the same
scale $Q=M$, the corresponding expressions of $V^{2LL}$ in the high 
and low energy regions read
\bea
V_{high}^{2LL}(\phi)&=&\rho[Q(\phi)]+(\beta_h-\beta_l)
\ln{M(\phi)\over m(\phi)}+\ {1\over 2}
\left(\cbeta_h-2\cbeta_l+\dot\beta_l\right)
\left[\ln{M(\phi)\over m(\phi)}\right]^2\ ,\nonumber\\
V_{low}^{2LL}(\phi)&=&\rho[Q(\phi)]
+\left.(\beta_h-\beta_l)\right|_M\ln{M(\phi)\over M}
+\ {1\over 2} \left.\left(\cbeta_h-2\cbeta_l+\dot\beta_l\right)\right|_M
\left[\ln{M(\phi)\over M}\right]^2\nonumber\\
&+&\left.  \left(\cbeta_l-\dot\beta_l\right)\right|_M\ln{m(\phi)\over 
M}\ln{M(\phi)\over M}\ .
\label{VHVL2}
\eea
The $V'$, $V''$ derivatives are straightforward to evaluate by taking into 
account
the $\phi-$dependence of the scale, i.e. $Q^2(\phi)=m^2+c^2\phi^2$. 
An alternative procedure, starting with the 1-loop initial potential
(\ref{Vfull1}), to incorporate the leading-log corrections at
arbitrary order is to perform the decoupling at the $\phi-$dependent
scale $M(\phi)$ (which leaves no threshold corrections)
and then evaluate the potential at the $m(\phi)$ scale. In this way 
one obtains expressions for the potential in the high and low energy
regions which coincide with those of the previous procedure. In particular,
collecting the 2-loop leading-log contributions one exactly recovers 
the result (\ref{VHVL2}).

To illustrate this general scheme of constructing the effective potential
across thresholds of new particles, let us consider again the simple
hybrid inflation model of sect.~3, but now with $N$ additional 
pairs of $\{H_+, H_-\}$ fields with invariant mass $M$. 
The superpotential of the full (high-energy) theory reads
\bea
\label{Wth}
W=\Phi\left(\sum_{a=1}^{N+1}\lambda H_+^aH_-^a-\mu^2\right)\ +\ 
\sum_{a=2}^{N+1} M H_+^aH_-^a
\ ,
\eea
where, for simplicity, we have taken equal Yukawa couplings for all the
$\{H_+, H_-\}$ pairs. For large enough $|\Phi|$ the tree-level
potential is flat, having the same form as in eq.~(\ref{rhoBD})
\bea
\label{rhoBDth}
V_0\equiv\rho=\mu^2 + {1\over 2}g^2\xi_D^2\ .
\eea
The 1-loop effective potential has the form (\ref{Vfull1}) (up to 
suppressed corrections) with
\bea
\label{BDmassesth}
m^2(\phi)={1\over 2}\lambda^2 \phi^2\;,\;\;\;\;\;
M^2(\phi)= M^2+{1\over 2}\lambda^2 \phi^2,
\eea
where $\phi=\sqrt{2} {\rm Re}\ \Phi$ and
\bea
\label{betaBDth}
\beta_l={1\over 8\pi^2}\Delta^2\;,\;\;\;\;\;
\beta_h={1\over 8\pi^2}(N+1)\Delta^2\ ,
\eea
with $\Delta=\sqrt{g^4\xi_D^2+\lambda^2 \mu^4}$.

Using the above RGEs, we can also include the two-loop 
leading log corrections, simply making use of the 
general 
expressions (\ref{Vfull1}) and (\ref{Vfull2}):
\bea
V(\phi) &= &\rho(Q) + {1\over 8\pi^2}\Delta^2(L_m +N L_M )
\nonumber\\
&+&{1\over (8\pi^2)^2} 
\left\{\lambda^4\mu^4\left[2L_m^2+2 N L_M 
L_m+N(N+1)L_M^2\right]-2\lambda^2g^2\mu^4(L_m^2+N L_M^2) \right.
\nonumber\\
&&\left.\;\;\;\;\;\;\;\;\;\;\;\;\;
+\ g^6\xi_D^2(2L_m^2+NL_mL_M+N^2L_M^2)
\right\}\ ,
\label{V2LLmodel}
\eea
where we have used the short-hand notation $L_m=\ln[m(\phi)/Q]$ and
$L_M=\ln[M(\phi)/Q]$. We have checked that this potential agrees with the 
one obtained by using the expressions for the two-loop potential 
of generic supersymetric models given in ref.~\cite{Martin} (which can 
also be used to add subleading logs and finite corrections).

The expressions for the potential above and below the threshold at $M$ can 
be easily obtained following  the effective
approach described above.
 More precisely, one can write
the potential in the high- and low-energy regions up to 1-loop or 
2-loop leading log order using eqs.~(\ref{VHVL}) and (\ref{VHVL2}) 
respectively. The $\dot\beta_{l,h}$ and $\cbeta_{l,h}$ derivatives 
for the latter can
be easily extracted from (\ref{V2LLmodel}) [by comparison
with (\ref{Vfull2})], or
calculated using the $\beta$-functions for the various 
parameters of the model involved in the expressions (\ref{betaBDth}),
which are given by eqs.~(\ref{betae}--\ref{betagv}) for the 
low-energy region and by eqs.~(\ref{betaeN}--\ref{betagvN}) for the
high-energy one. 
We do not present explicitly the corresponding formulas. 

The results are somewhat better than in the case without thresholds 
analyzed in sect.~3. Nevertheless the conclusions are basically the same: 
The number of e-folds and the slope of the spectral index are correlated 
by an equation similar to (\ref{rule}), so that it is not possible to get 
$dn/d\ln k={\cal O}(-0.05)$ and $N_e=50-60$ at the same time. It is worth 
remarking that this happens in spite of the fact that the $\beta$ 
functions undergo quick changes in the threshold region, as desired. 
However, the corrected potential [{\it e.g.} as given in 
eq.~(\ref{VHVL2})] softens 
these effects and the net impact in $N_e$ and $dn/d\ln k$ gets much 
reduced.

\subsection{Thresholds of new physics above the inflationary scale}

We consider now an even simpler modification of the inflationary
potential. It assumes a scale of new physics, $M$, higher than the scales 
relevant to inflation. Nevertheless, inflation is affected through the 
presence of non-renormalizable operators of the inflaton field that lift 
the flat direction along which the inflaton rolls. It could appear naively 
that such operators should have a negligible impact on inflation whenever 
they can be reliably taken into account ({\it i.e.} $\phi^2/M^2\ll 1$) but 
we show in this section that this is not the case. In fact, through such 
effects we are able to give a simple and complete potential that satisfies 
all three goals [{\it i)-iii)}] listed in section~3. The analysis can
be carried out in a very model-independent way.

Let us start directly with an inflaton potential that reads
\be
V(\phi)=\rho+\beta \ln{m(\phi)\over Q}+\phi^4{\phi^{2N}\over M^{2N}}\ 
.
\label{VNRO}
\ee
The first two terms just correspond to the generic one-loop potential we 
have discussed in previous sections. In the 
small-coupling regime, as in subsect.~2.1, we 
take $\beta$ as a constant. The last term in 
(\ref{VNRO}) is a non-renormalizable operator (NRO) left in the low-energy 
theory after integrating out some unspecified physics at the high scale 
$M$. This scale absorbs any possible coupling in front of the operator. 
Of course $V(\phi)$ may contain other NROs of different order. Here we 
assume that the one shown in eq.~(\ref{VNRO}) is the lowest order one; 
NROs of higher order will have a negligible impact compared to it 
(provided they are suppressed by the same mass scale $M$).
The sign and power we have assumed for this NRO are convenient to 
guarantee the stability of the potential. Notice also that an even power 
for this operator is what one expects generically in supersymmetric theories. 
In fact, the question of whether such potential can have a supersymmetric 
origin is very interesting and we leave its discussion for the end of this 
section. 

Now we focus on studying whether a potential of the form 
(\ref{VNRO}) can reproduce the WMAP indications for a running $n$, and for 
what values of the potential parameters can it do so. By trivial 
inspection of the derivatives of $V(\phi)$ with respect to $\phi$:
\bea
V'(\phi)&=&{\beta\over\phi}+2(N+2)\phi^3{\phi^{2N}\over 
M^{2N}}\ ,\nonumber\\
V''(\phi)&=&-{\beta\over\phi^2}+2(N+2)(2N+3)\phi^2{\phi^{2N}\over 
M^{2N}}\ ,\nonumber\\
V'''(\phi)&=&2{\beta\over\phi^3}+4(N+2)(2N+3)(N+1)\phi{\phi^{2N}\over 
M^{2N}}\ ,
\label{derivatives}
\eea
we realize that the NRO can have a significant impact on inflation when the 
small number $(\phi/M)^{2N}$ is comparable in size to $\beta/\phi^4$ 
(which is also quite small). In other words, even for $\phi^2\ll M^2$, the 
NRO can have an important effect because it gives a correction to a 
potential which is almost flat.

It is also immediate to realize from (\ref{derivatives}) that, for 
sufficiently large $\phi$, the higher derivatives
$V'', V'''$ (and thus $\eta, \xi$) can receive a large 
contribution  from the NRO while the contribution to $V'$ (and 
thus $\epsilon$) is much less significant,
thanks to the additional $(2N+3)$ and $(2N+3)(2N+2)$
factors in $V'', V'''$. In consequence,
as $\phi$ rolls down its potential the 
effect of the NRO quickly dies away and inflation proceeds as discussed in 
section~2 but, in the early stages of inflation the 
running of $n$ can get important modifications from the NRO corrections,
even though the values of $V$ and (to some extent) $V'$ are scarcely 
modified. Let us examine this in more 
quantitative terms.

The slow-roll parameters (\ref{SlowRoll}) can be readily found as a function 
of $\phi$ from 
the derivatives of the potential listed in (\ref{derivatives}). The 
$\epsilon$ parameter is much smaller than $\eta$ and $\xi$, so  
$n\simeq 1+2\eta$ and $d n/d \ln k\simeq -2\xi$. In this way we obtain 
\be
{dn\over d\ln k}\simeq -{1\over 
N_e^2(\phi)}\left[1+(2N+3)(N+1)r(\phi)\right][1+r(\phi)]\ ,
\label{dndlnkphi}
\ee
where 
\be
r(\phi)\equiv 2(N+2){\phi^4\over\beta}{\phi^{2N}\over M^{2N}}\ ,
\ee
and 
\be
N_e(\phi)= {\rho\phi^2\over 2\beta M_p^2}\ ,
\label{Nephi}
\ee
which is approximately the number of e-folds from $\phi$ till the end of 
inflation [obtained by neglecting the NRO contribution to $V'(\phi)$ 
in eq.~(\ref{Ne})]. Applying eq.~(\ref{dndlnkphi}) to the starting point 
$\phi_*$, it looks possible to get $N_e^2 (dn/d\ln k|_*)\sim -125$, as 
suggested by WMAP, if $N$ and/or $r(\phi_*)$ are not small. We analyze 
next whether this is feasible.

Let us call $\phi_0$ the value of the 
inflaton at that particular point with $n=1$. At $\phi_0$ we should have 
$V''(\phi_0)\simeq 0$, which implies $r(\phi_0)\simeq 1/(2N+3)$. 
Using the approximation of eq.~(\ref{Nephi}) it is simple to get
\be
r(\phi_*)\simeq {1\over 2N+3}\left({N_e\over N_e^0}\right)^{N+2}\ ,
\label{rstar}
\ee
where $N_e=N_e(\phi_*)$ is the total number of e-folds, and 
$N_e^0=N_e(\phi_0)$; {\it i.e.} $n=1$ after the first $N_e-N_e^0$ e-folds.
\begin{figure}[t]
\vspace{1.cm} \centerline{
\psfig{figure=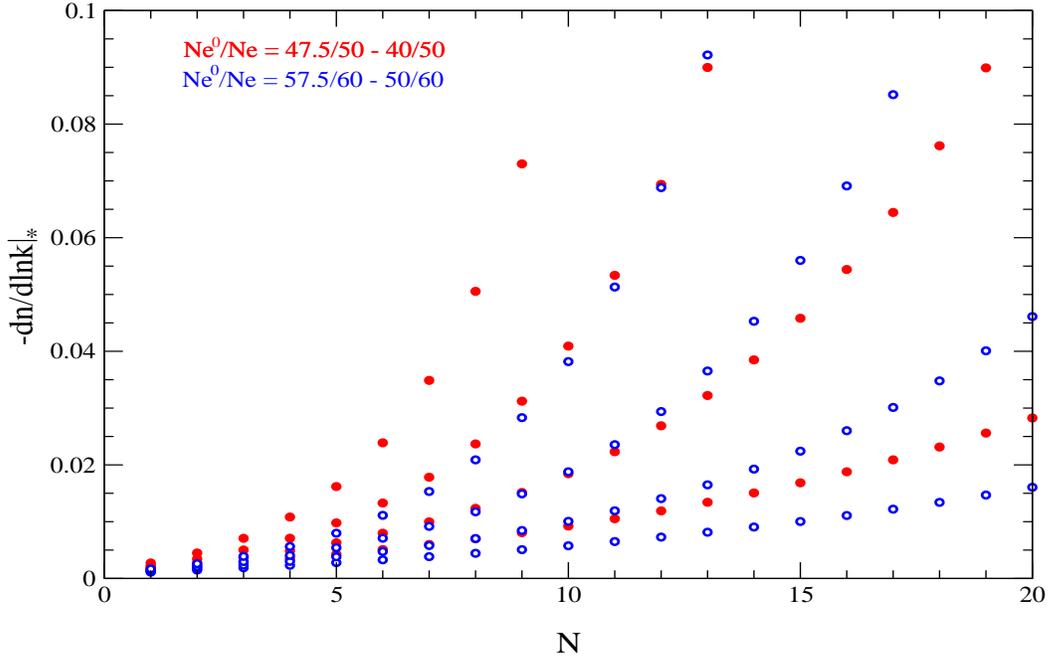,angle=-90,height=8cm,width=8cm,bbllx=4.cm,%
bblly=7.cm,bburx=20.cm,bbury=21.cm}}
\caption{\footnotesize
Running of the spectral index, given by $-dn/d\ln k|_*$, as a 
function of $N$ for several values of $N_e^0/N_e$. Red solid (open blue) 
dots have $N_e=50$ $(60)$ and $N_e^0=\{47.5, 45, 42.5, 40\}$ ($\{57.5, 55, 
52.5, 50\}$), with higher values of $N_e^0$ giving smaller running.} 
\label{fig:dndlnk}
\end{figure}
Using Eqs.~(\ref{dndlnkphi}-\ref{rstar}) we can find out what 
values of $N$ are required to get enough running for $n$. This is shown in 
Figure~\ref{fig:dndlnk}, which gives $dn/d\ln k|_*$ as a function of $N$ 
for 
different 
choices of $N_e^0/N_e$ [corresponding to $k(\phi_0)$ not far from 
$10^{-2} {\rm Mpc}^{-1}$]. From this plot we see that large values of $N$ 
are required to get a significant running of $n$, with smaller $N_e^0/N_e$ 
being preferred. As an example, let us take $N=9$ and $N_e^0/N_e=42.5/50$, 
which gives $dn/d\ln k|_*\simeq -0.03$.

\begin{figure}[t]
\vspace{1.cm} \centerline{
\psfig{figure=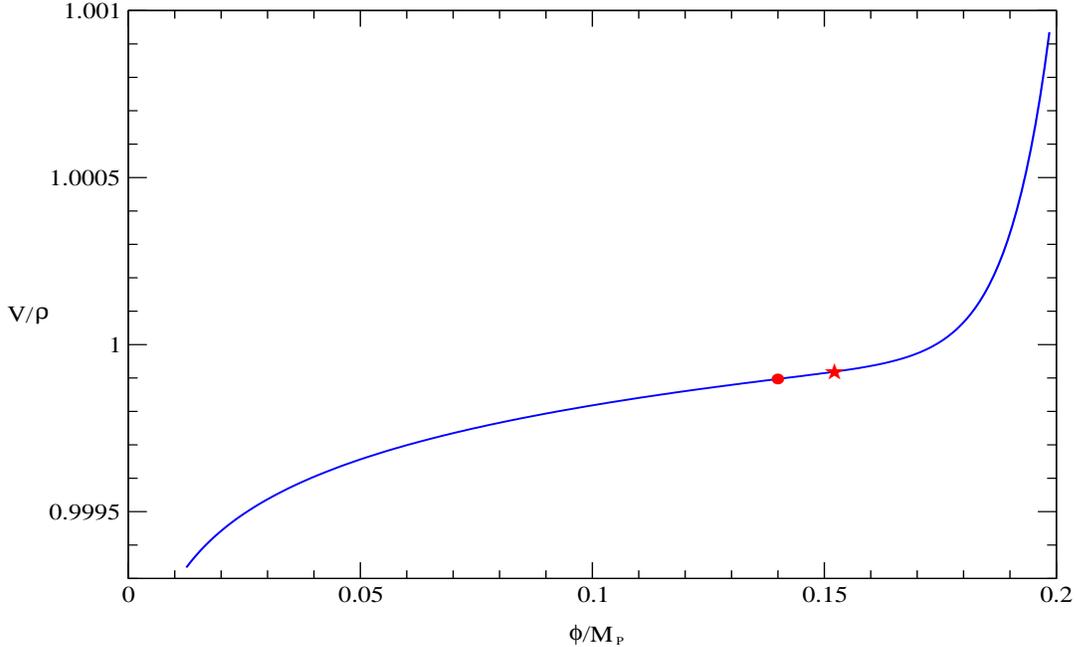,angle=-90,height=8cm,width=8cm,bbllx=4.cm,%
bblly=7.cm,bburx=20.cm,bbury=21.cm}}
\caption{\footnotesize Inflaton effective potential (normalized to $\rho$) 
with a NRO as in eq.~(\ref{VNRO}) for $N=9$ and parameters as given in 
the text. The star marks $\phi_*$ and the circle, $\phi_0$.
} 
\label{fig:VNRO}
\end{figure}

\begin{figure}[tbp]
\vspace{1.cm} \centerline{
\psfig{figure=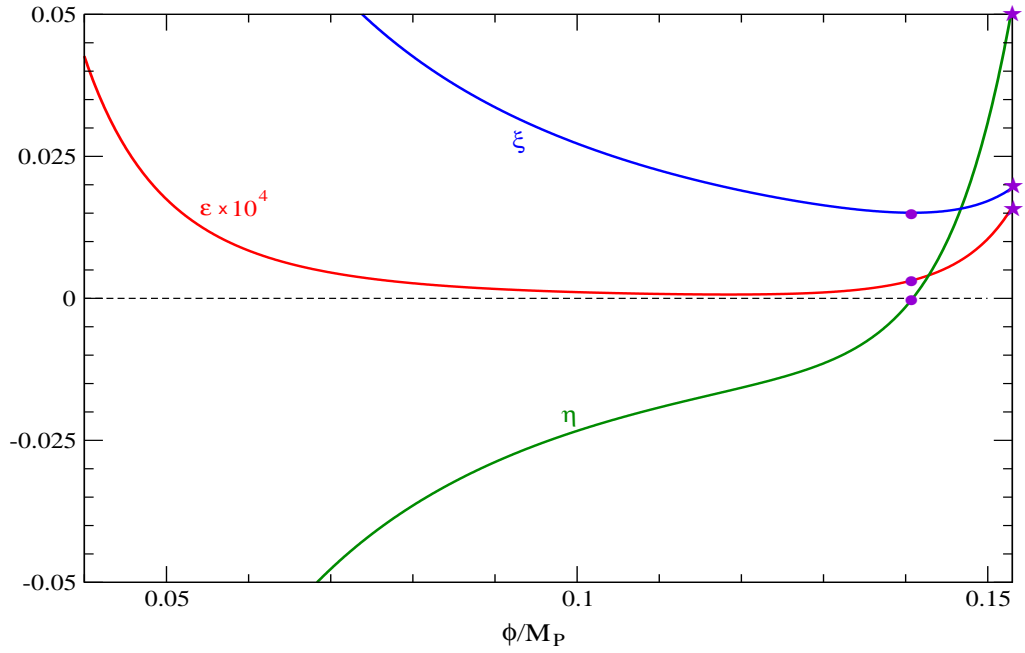,angle=-90,height=8cm,width=8cm,bbllx=4.cm,%
bblly=7.cm,bburx=20.cm,bbury=21.cm}}
\vspace*{2cm}
\centerline{
\psfig{figure=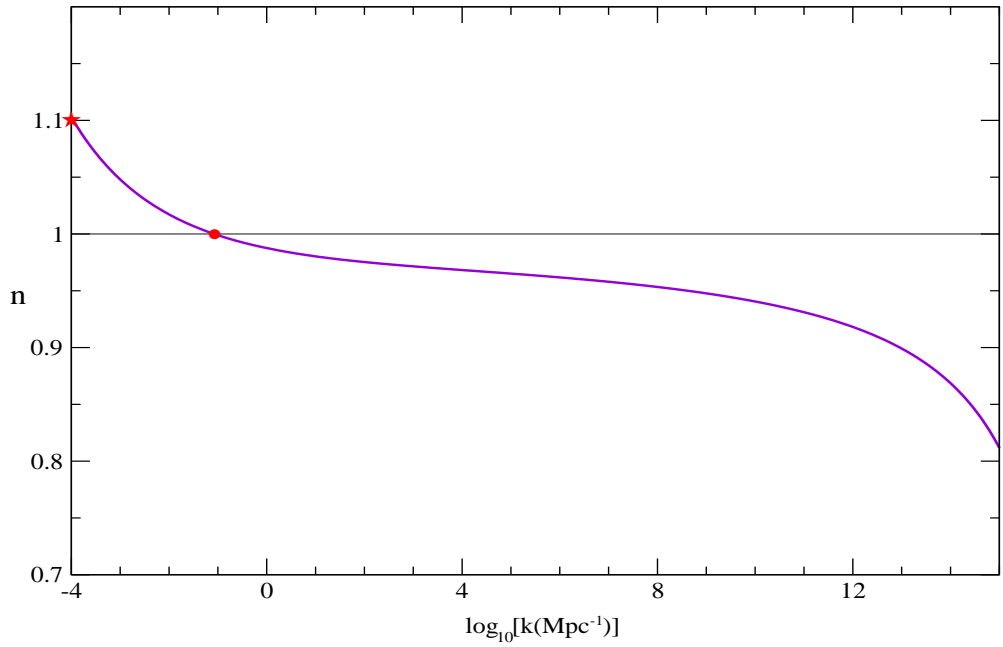,angle=-90,height=8cm,width=8cm,bbllx=4.cm,%
bblly=7.cm,bburx=20.cm,bbury=21.cm}}
\caption{\footnotesize Slow-roll parameters (upper plot) and 
scalar spectral index as a function of scale in ${\rm Mpc}^{-1}$
(lower plot) for the inflaton potential of Figure \ref{fig:VNRO}. }
\label{fig:slownNRO}
\end{figure}

For a fixed value of the scale of new physics, $M$, the quantities 
$\beta$ and $\rho$ are determined 
by the constraints of $P_k$ and $N_e$:
one finds $\rho\simeq (10^{-3}\sqrt{M M_p})^4$ and $\beta\simeq 
(10^{-4}M)^4$. Choosing for instance $M\simeq 0.95M_p$ one further gets 
$dn/d\ln k|_*\simeq -0.03$ for $\phi_*\simeq 0.15 M_p$ while 
$\phi_0\simeq 0.142 M_p$; the numerical values for the 
number of e-folds are $N_e\simeq 48.8$ and $N_e^0\simeq 42.4$; and one  
gets $P_k\simeq (2.95\times 10^{-9})\times(0.8)$. Note that 
$\phi_*/M$ is sufficiently small for the effective theory with the NRO to 
be trustable. Figure~\ref{fig:VNRO} shows the effective potential as a 
function of $\phi/M_p$, with $\phi_*$ indicated by a star and $\phi_0$ by 
a circle. Notice how $\phi_*$ is below the range where the NRO starts 
to be important for $V'(\phi)$ [but not for $V''(\phi)$]. The upper plot 
in Figure~\ref{fig:slownNRO} shows the slow-roll parameters as a function 
of $\phi/M_p$. Inflation does not continue below $\phi\simeq 0.02 M_p$. 
Finally, the lower plot of Figure~\ref{fig:slownNRO} gives the scalar 
spectral index as a function of $\log_{10}k ({\rm Mpc}^{-1})$.

The running of the spectral index with $k$, as shown in the 
picture just mentioned, can be well approximated by 
\bea
\label{nintegratedN}
n & = & n_* + {1\over N_e} - {1\over N_e - \ln (k/ k_*)}\nonumber\\
&-&{N_e\over (N+1)}\left(\left.{d n\over d\ln k}\right|_*+{1\over N_e^2} 
\right)\left[\left(1- {1\over N_e}\ln{k\over k_*}
\right)^{N+1}-1\right]\ .
\eea
The first line is the same as eq.~(\ref{nintegrated}), which is valid for  
(weak coupling) potentials with a radiative dependence on the logarithm of 
the inflaton field. The effect of the NRO is explicitly seen  as the 
additive term in the second line.

As already discussed in connection with Figure~\ref{fig:dndlnk}, stronger 
running of the spectral index can be achieved if $N$ is larger. This 
leads us to the next important question: how reasonable is it to expect a 
non-renormalizable operator with {\it e.g.} $N=9$ or even larger? Recall 
that, for the previous mechanism to work, such operator should be the 
leading one. This is not a serious problem because supersymmetric flat 
directions as the one we are using for inflation can be protected against 
lifting by additional symmetries. In fact it is common that some of the 
flat directions of supersymmetric models are only lifted by NROs at very 
high order. This is well known, for instance in the MSSM \cite{MSSMflat}. 
This is also very common in the context of D=4 string theories, where
string selection rules forbid many operators in the superpotential that
would be allowed just by gauge invariance. For example, in the popular 
$Z_3-$orbifold compactifications of the heterotic string, 
non-renormalizable couplings involving twisted matter fields
(which would be the relevant case if the inflaton is a twisted
field) have the structure $\Phi^{3+9n}$, with $n=1,2, ...$
\cite{NROstring}.

Concerning the supersymmetric realization of a potential like that in 
(\ref{VNRO}) the first guess would be to use a superpotential of the form
\be
W=\Phi(\lambda H_+H_- -\mu^2)+{1\over (N+3)}\Phi^3{\Phi^N\over M^N}\ ,
\label{WNRO}
\ee
which is simply the standard superpotential we used in section~3 
supplemented by a non-renormalizable term for the inflaton. From 
(\ref{WNRO}), one obtains at tree level the inflaton potential
\be
V=V_D+\left|\Phi^2{\Phi^N\over M^N}-\mu^2\right|^2\ ,
\label{VWNRO}
\ee
where $V_D$ is the Fayet-Iliopoulos contribution. This potential
has a term of order $\Phi^4\Phi^{2N}/M^{2N}$ like the one we are 
after, but also a term of lower order $(-\mu^2\Phi^2{\Phi^N/M^N}
+{\rm h.c.})$. For the higher order term to dominate one would need
$|\mu^2|\ll \Phi^2{\Phi^N/M^N}$ which can in principle be arranged. 
In any case, the presence of a non-zero $\mu$ poses a different problem. 
In view of the form of the potential (\ref{VWNRO}) it is clear that there 
will be a series of minima with $\Phi=\Phi_m \exp[i\pi/(2+N)]$ where 
$\Phi_m^{(N+2)}=|\mu^2|M^N$. The flat potential for $\Phi$ is therefore 
lifted by the NRO along some directions in the complex plane of $\Phi$ 
while along other directions the potential develops minima and this can 
change qualitatively the evolution of the inflaton. It is interesting 
that the usual logarithmic one-loop corrections that cause the 
inflaton to roll in the standard scenario can also cure the previous 
problem in the present case so that the minima are no longer dangerous.
In that case one can have the inflaton running 
in a stable trajectory along which only its real part is non-zero.
Nevertheless, the simplest cure of the previous problems is to choose 
$\mu=0$ which, incidentally, is the choice of the original formulation of 
the $D$-term inflation model \cite{BD}. In that case one simply gets a 
potential term of 
order $\phi^4\phi^{2N}/M^{2N}$ as in eq.~(\ref{VNRO}). Notice how, thanks 
to supersymmetry, the order of the NRO in the superpotential (\ref{WNRO}) 
has almost doubled in the potential (\ref{VWNRO}).

An appealing alternative is to start with 
\be
W=\lambda\Phi H_+H_- + {1\over 2}m 
{\Phi'}^2+{1\over (P+2)}\Phi'\Phi^2{\Phi^P\over M^P}\ 
,
\label{WNRO2}
\ee
where we have introduced an extra field $\Phi'$ with mass $m\ll M$ (so 
that $\Phi'$ really belongs in the effective theory below $M$). When one 
is along the flat direction for the inflaton field $\Phi$ the coupling to 
$\Phi'$ generates a tadpole for it so that also $\Phi'$ develops a VEV. As 
we are only interested in the evolution of the inflaton field, we can 
simply eliminate $\Phi'$ by solving its equation of motion in terms 
of $\Phi$ and substituting in the two-field potential $V(\Phi,\Phi')$ 
to get the final potential for $\Phi$: $V(\Phi)=V(\Phi,\Phi'(\Phi))$. In 
this way one gets the following inflaton potential (along $\phi=\sqrt{2}\ 
{\rm Re}\ \Phi$ and absorbing factors of $\sqrt{2}$ in $M$)
\be
V=V_D+\phi^4{\phi^2\over m^2}{\phi^{4P}\over M^{4P}}\ ,
\ee
where again $V_D$ is the Fayet-Iliopoulos contribution. The advantage of 
this option is that the NROs are naturally of higher order. For instance, 
to get $N=9$ one needs a rather modest $P=4$ in the superpotential 
(\ref{WNRO2}).

\section{Conclusions}

In this paper we have assumed that the WMAP indications on the running of 
the scalar spectral index are real and then explored the intriguing 
possibility that this runnig is caused by physics at very high energy 
scales affecting the inflaton potential. It is tantalizing to imagine that 
we could be seeing in the sky the imprints of such physics. In more 
concrete terms we aim for well motivated inflaton potentials (from the 
point of view of particle physics) that can reproduce the suggested slope 
of $n$ with scale, $dn/d\ln k\sim {\cal O}(-0.05)$, giving at the same 
time the right amount of e-folds and the right amplitude of the power 
spectrum of scalar fluctuations.

This is not an easy task and has atracted some effort recently. We have 
focused on a large class of inflationary models which we find 
particularly atractive: those with flat tree-level potential lifted by 
radiative corrections. These radiative corrections directly control the 
slow roll of the inflaton and, in turn, the running of the spectral index 
$n$. These models include some typical hybrid inflation models as a 
particularly interesting subclass. In these models there is a one-to-one 
correspondence between the value of the inflaton potential, the related 
distance scale $k$ and the associated renormalization scale, so that the 
running of the inflaton down its potential scans different high energy 
scales and cosmological distances. In this language, WMAP indicates that 
the spectrum of scalar perturbations is blue ($n>1$) at the largest 
cosmological scales ($k\sim 10^{-4}{\rm Mpc}^{-1}$) and turns to red 
($n<1$) 
below some scale $k\sim 10^{-2}{\rm Mpc}^{-1}$. In terms of energy 
scales, at the start 
of inflation the inflaton had a very large value (corresponding to very 
high energy scales) and its slow-roll produced $n>1$ and a negative slope 
$dn/d\ln k$. After a few e-folds this negative slope made $n$ cross 1 at 
some particular value $\phi_0$ of the inflaton field, corresponding to 
some particular energy scale $Q_0$. Later on, $n$ gets stabilized around 
$0.94-0.99$ and the additional number of e-folds accumulates till 
$N_e\sim 50-60$, when inflation ends.

In our search for suitable potentials in this type of models we find two 
different regimes depending on how sizeable the coupling constants of 
the model parameters are. In 
the small-coupling regime, these models make sharp predictions for the 
size of the spectral index $n$ and for its running with scale, namely
\be
- {dn\over d\ln k}=(n-1)^2\ll 1\ .
\ee 
From this equation, it is clear that the preliminary WMAP indication of 
$n$ crossing $n=1$ cannot be reproduced. 

On the other hand, at larger coupling we find that a much stronger
running of $n$ is 
absolutely natural and even unavoidable. However, reproducing the WMAP 
indication for $dn/d\ln k$ {\em and} a sufficient number of e-folds ($N_e\sim 
50-60$) simultaneously is very difficult. We have illustrated this
in a popular $D$-term hybrid inflation model for which we have obtained a 
`no-go' constraint of the form (\ref{rule})
\be
(N_e^0)^2 \left.{dn\over d\ln k}\right|_{Q_0}\simeq -1.1\ ,
\label{concl2}
\ee
where, as mentioned above, $Q_0$ is the scale at which $n$ crosses 1, and 
$N_e^0$ is the number of e-folds from that scale till the end of inflation.
Now, if one requires $dn/d\ln k\simeq -0.05$, then $N_e^0\ll 50$.
Therefore, unless a subsequent stage of inflation completes the total 
number of e-folds, these models are in trouble by themselves to reproduce 
the behaviour of $n$ mentioned above.

As the details of inflation in this region of scales are sensitive to 
possible physics thresholds, we considered the natural possibility of 
the inflaton coupling to heavy new fields with a mass $M$ that falls in 
the energy range crossed by the inflaton field in the first few e-folds. 
Somewhat to our surprise we found that, in the presence of sharp 
thresholds, the results of the previous case are not changed 
substantially and still one finds that the requirement of a significant 
running of $n$ leads to a small number of e-folds being produced.

A different way in which high energy physics can leave an imprint in the 
predictions of inflation is through non-renormalizable operators that 
modify the inflaton potential in the inflationary energy range . In this 
case, the new energy  threshold is not crossed during inflation (this is 
necessary for a reliable treatment of new physics effects in terms of 
effective operators in the first place!). The key point here is that even 
operators that are suppressed by a high energy mass scale can become 
important in the environment of a very flat potential. In this context we 
are able to write down very simple and well motivated inflationary 
potentials that can give a running spectral index in agreement with WMAP 
and a sufficient number of e-folds for reasonable choices of 
the parameters. The succesful type of potential combines three 
ingredients: tree-level flatness, smooth dependence on the 
inflaton through radiative logarithms and a non-renormalizable term:
\be
V(\phi)=\rho+\beta \ln{m(\phi)\over Q}+\phi^4{\phi^{2N}\over M^{2N}}\ .
\label{NROconcl}
\ee
We have further motivated this type of potentials by finding 
supersymmetric realizations through superpotentials that include 
non-renormalizable operators (NROs) of 
modest order.

Finally, we summarize the functional dependence of $n(k)$ expected for 
flat tree-level potentials. In the small-coupling regime, and if no
mass thresholds are crossed during inflation,
this is simply given by
\bea
\label{smallrunning2concl}
n(k) =1 -{1\over N_e - \ln (k/ k_*)} 
\ ,
\eea
where $k_*$ is the initial distance scale 
$k_*\sim 10^{-4}\ {\rm Mpc}^{-1}$. 
The only parameter in (\ref{smallrunning2concl}), $N_e$,
corresponds typically to the total number of e-folds,
i.e. $N_e\simeq 50-60$, although it could be either
smaller [if there are subsequent episodes of inflation], or larger
--or even negative--
[if the end of inflation is not marked by $\eta={\cal O}(1)$,
but by the violation of some hybrid-inflation condition for 
the tree-level flatness]. 
%
This is explained in subsect.~2.1.
%
%
The case with thresholds and/or not-so-small couplings are
addressed in subsects.~2.1,~2.2,~4.1.

On the other hand, when there are effects of high-energy (new) 
physics through non-renormalizable operators (NROs), $n(k)$ is of the form
\bea
\label{nintegratedNconcl}
n (k)& = & n_* + {1\over N_e} - {1\over N_e - \ln (k/ k_*)}\nonumber\\
&-&{N_e\over (N+1)}\left(\left.{d n\over d\ln k}\right|_*+{1\over N_e^2} 
\right)\left[\left(1- {1\over N_e}\ln{k\over k_*}
\right)^{N+1}-1\right]\ ,
\eea
where $*$ denotes quantities evaluated at the initial scale $k_*$, 
and $N$ is the exponent of the dominant NRO, as expressed in 
eq.~(\ref{NROconcl}).
The quantity $N_e$ has the same meaning as in 
eq.~(\ref{smallrunning2concl}). The other parameters in
eq.~(\ref{nintegratedNconcl}) are $n_*$, $d n/d\ln k |_*$,
which can be expressed in terms of the parameters of the
inflaton potential (\ref{NROconcl}), as explained in 
subsect.~4.2. 
Finally, if the effect of the NROs
is negligible [e.g. if $M\rightarrow\infty$ in 
(\ref{NROconcl}), which corresponds to $n_*\rightarrow 1-1/N_e$,
$d n/d\ln k |_*=-1/ N_e^2$ in~(\ref{nintegratedNconcl})], then 
eq.~(\ref{smallrunning2concl}) is recovered. For a more detailed
discussion, see subsect.~4.2.

It would be a very nice test for the wide and well-motivated class
of inflation models analyzed in this paper, 
if future analyses and measurements of 
the spectral index $n(k)$ could be adjusted by one of the
previous expressions, which in turn would provide precious
information about the inflaton potential
and on  physics at very high energy-scales.

\subsection*{A. The simple hybrid-inflation model with N+1 flavours}
\setcounter{equation}{0}
\renewcommand{\theequation}{A.\arabic{equation}}

We consider here the same hybrid-inflation model of sect.~3, with
$N+1$ pairs of $H_\pm$ fields, instead of the unique pair (i.e. $N=0$) of the
original model. The results have interest on their own (e.g. they
illustrate how the values of the coupling constants and the speed
of running are affected by the number of flavours), but they are
also useful to illustrate the threshold-crossing procedure discussed
in sect.~4.1.

The superpotential reads now
\bea
\label{WN}
W=\Phi(\sum_{a=1}^{N+1}\lambda_a H_+^aH_-^a-\mu^2)\ ,
\eea
where $\lambda_a$ are the $N+1$ Yukawa couplings. As for the $N=0$ case,
$H_\pm^a$ have charges $\pm 1$ with respect to the $U(1)$ gauge group
with gauge coupling $g$ and Fayet-Iliopoulos term $\xi_D$. The 
tree-level scalar potential is given by $V_0=V_F +V_D$,
with
\bea
\label{VFVDN}
V_F&=&\left|\sum_{a}\lambda_a H_+^aH_-^a-\mu^2\right|^2 
+ \sum_{a}\lambda_a^2\left(\left| H_-\right|^2 
+ \left|H_+\right|^2\right)\left|\Phi\right|^2\ ,\nonumber\\
V_D&=&{g^2\over 2}\left[\sum_{a}\left(\left|H_+^a\right|^2 
- \left|H_-^a\right|^2 \right)
+ \xi_D\right]^2\ .
\eea
Again, the global minimum of the potential is supersymmetric
($V_F=0$, $V_D=0$), but for large enough $|\Phi|$
the potential has a minimum at $H_\pm^a=0$ and
is flat in $\phi$. The tree-level potential in this region is 
\bea
\label{rhoBDN}
\rho=\mu^4 + {1\over 2}g^2\xi_D^2\ .
\eea
The various $\beta$-functions, defined as derivatives with respect to $\ln Q$, 
are given by
\bea
\label{betaeN}
\beta_g&=&{1\over 8\pi^2}(N+1) g^3\ ,\\
\label{betaxiN}
\beta_{\xi_D}&=& 0\ ,\\
\label{betagN}
\beta_{\lambda_a}&=&{1\over 16\pi^2}\lambda_a\left[3\lambda_a^2 
+\sum_{b\neq a} \lambda_b^2 - 4g^2\right]\ ,\\
\label{betagvN}
\beta_{\mu^2}&=&{1\over 16\pi^2} \sum_{a=1}^{N+1} \lambda_a^2 \mu^2 \ ,\\
\label{betaBDN}
\beta_\rho&=&{1\over 8\pi^2}\left[(N+1)g^4\xi_D^2+\sum_{a=1}^{N+1} 
\lambda_a^2 
\mu^4\right]\ .
\eea

For simplicity we can take the same
Yukawa coupling $\lambda_a=\lambda$ for all $\{H_+^a, H_-^a\}$ pairs, 
which is a stable condition under RG evolution.

\section*{Acknowledgements}
We thank Alejandro Ibarra for very interesting suggestions and Toni Riotto for 
a careful reading of the manuscript. This work is 
supported by the Spanish Ministry of Education and Science through a 
M.E.C. project (FPA2004-02015). Guillermo Ballesteros acknowledges the 
financial support of the Comunidad de Madrid and the European Social Fund 
through a research grant.

\end{document}